\magnification\magstep1
  \def\aa{\vskip 0.5 true cm}
  \parskip = 0.4 true cm
  \pageno = 0
  \baselineskip = 0.85 true cm
  \vsize = 22 true cm
  \hsize = 16 true cm
  \def\sa{\vskip 0.30 true cm}
  \def\sb{\vskip 0.60 true cm}

\baselineskip = 0.6 true cm

\rightline{\bf LYCEN 9138}
\rightline{22 July 1991}

\sb
\sb

\centerline {\bf Intensity of two-photon absorption transitions 
for Ni$^{2+}$ in MgO}

\sa
\sb

\centerline {J.~Sztucki$^*$}

\sa

\centerline {Institut f\"ur Festk\"orperphysik,}
\centerline {Technische Hochschule Darmstadt,} 
\centerline {6100 Darmstadt, Germany}

\sa
\sa 

\centerline { M.~Daoud and M.~Kibler } 

\sa

\centerline {Institut de Physique Nucl\'eaire de Lyon,}  
\centerline {Institut National de Physique Nucl\'eaire}
\centerline {et de Physique des Particules~--~Centre}
\centerline {National de la Recherche Scientifique et}
\centerline {Universit\'e Claude Bernard,}
\centerline {F-69622 Villeurbanne Cedex, France}         

\sa
\sb

\baselineskip = 0.5 true cm

\sa

\centerline {\bf Abstract}

\sa

The parity-allowed two-photon transitions between the ground state 
$^3A_2(T_2)$ of the configuration $3d^8$ in cubical symmetry and the 
excited states of the same configuration are obtained via a simple 
model. This model is developed in a symmetry adapted framework by using 
second-order mechanisms and ionic wave-functions.   It is 
applied  to  the  recent  experimental results obtained by McClure 
and co-workers for Ni$^{2+}$ in MgO.

\sb
\sb
\sb
\sa
\sa

\noindent Published in the Physical Review B 
          ({\bf Phys.~Rev.~B 45 (1992) 2023-2028}). 

\noindent $^*$On leave of absence from the Institute for Low Temperature 
and Structure Research, Polish Academy of Sciences, 50~-~950 
Wroc\l aw, Poland.

\vfill\eject
\baselineskip = 0.85 true cm

\centerline {\bf I. INTRODUCTION}

In recent years, two-photon absorption spectroscopy of partly-filled shell 
(rare-earth and transition-metal) ions in solids have received a great deal 
of attention. Major achievements have been done both from an  
experimental and theoretical viewpoint (see, for instance, 
Refs.~1-13). The main theoretical developments concern~: (i) the use, 
in addition to the Axe second-order mechanism, of  
higher order mechanisms describing spin-orbit, crystal-field, 
ligand polarizibility, and electron correlation effects 
and (ii) the application of 
symmetry adaptation techniques. More specifically, it has been 
shown recently how quantitative, rather than only 
qualitative,$^2$ 
symmetry considerations can be used to obtain the polarization 
dependence of a two-photon absorption between two Stark (rather 
than $[^{2 S + 1}L]_J$) levels.$^{8,9,11}$

Recently, results on two-photon spectroscopy of Ni$^{2+}$ 
in MgO have been reported by Moncorg\'e and Benyattou$^{12}$ 
and by McClure and co-workers.$^{13}$ It 
is then the aim of the present paper to further contribute to test 
theoretical models for analysing two-photon spectroscopy by 
using the fresh data of Ref.~13. 

We shall develop, in Secs.~II and III, a simple 
quantitative model for dealing with the $d^8$ (or $d^2$) 
configuration in octahedral symmetry. The results provided by 
this model for MgO:Ni$^{2+}$ are given in Sec.~IV and 
discussed in Sec.~V. 
\aa

\centerline {\bf II. THEORY} 

The transition matrix element $M_{i \rightarrow f}$ for a two-photon absorption
between an initial state $i$ and a final state $f$ is given by the
G\"oppert-Mayer formula
$$
M_{i \rightarrow f} \; = \; \sum_v \;
  {{(f | {\vec D}. \, {\vec {\cal E}} |v) \; 
    (v | {\vec D}. \, {\vec {\cal E}} |i)}\over
   {\hbar \omega - E_v}}
\ , \eqno (1)
$$
where the sum on $v$ extends over all intermediate states. In Eq.~(1),
$E_v$ denotes the energy of the state $v$ with respect to the one of the state
$i$. Here, we consider two identical photons and use single mode excitations of
the radiation field with energy $\hbar \omega$, wave vector $\vec k$, and unit
polarization vector ${\vec {\cal E}}$. Equation (1) is derived in the framework
of the dipolar approximation and ${\vec D}$ refers to the dipole moment
operator for the eight electrons of the configuration $3d^8$ while 
${\vec D} . \, {\vec {\cal E}}$ stands for the scalar product 
of $\vec D$ and ${\vec {\cal E}}$.

The sum on $v$ in (1) can be handled by using a quasi-closure 
approximation.$^1$ As a result, 
$M_{i \rightarrow f}$ turns out to be the matrix
element of an effective operator $H_{eff}$ between the initial and final 
electronic state vectors.$^{1,3}$ Following the phenomenological argument of 
Ref.~8, the operator $H_{eff}$ may be extended to the form
$$
H_{eff} \; = \; \sum_{k_Sk_Lk} \;
  C \left [
  \big ( k_Sk_L \big )k\right ] \;
  \big ( {\bf W}^{(k_Sk_L)k} \, . \, \{{\cal E} {\cal E}\}^{(k)}
  \big )
\ . \eqno (2)
$$
The polarization dependence in Eq.~(2) is given by the tensor product 
$\{{\cal E} {\cal E}\}^{(k)}$ 
where $k$ can be only 0 and 2 for identical photons. The electronic dependence
is contained in the double tensor ${\bf W}^{(k_Sk_L)k}$ 
of spin rank $k_S$, orbital rank $k_L$, and
total rank $k$. The electronic and polarization parts are coupled together
through the scalar product (.) occuring in (2). The 
parameters $C\left [ (k_Sk_L)k\right ]$
depend on the configuration
$3d^8$ and on the $3d^7 n' \ell '$ and $n' {\ell '} ^{4 \ell ' + 1} 3d^9$
configurations to which the intermediate state vectors belong (the most
important excited configuration is probably $3d^74p$). These 
parameters are clearly model-dependent and may be calculated 
from first principles. For example, the parameter 
C[(02)2], which we shall use in Sec.~IV, reads 
$$
C \big [ (02)2 \big ] \; = \;
 - \; {\sqrt 2} \; e^2 \;
 \sum_{n'\ell '} \; 
 \left [\hbar \omega - E(n' \ell ')\right ]^{-1} \;
 \big( 3d | r | n' \ell ' \big) ^2 \; 
 \big(2 \Vert C^{(1)} \Vert \ell '\big)^2 \;
 \left\{
 \matrix{ 
 1 & 2 & 1 \cr
 2 & \ell '& 2\cr
 }\right\}
\ , \eqno (3)
$$
where $E(n'\ell ')$ is an average energy arising from the quasi-closure
approximation.

In Eq.~(2), the contribution $(k_S = 0, k_L = 2, k = 2)$ described by 
(3) corresponds to the second-order mechanism first introduced by 
Axe.$^{1}$ The other contributions 
$(k_S \ne 0, k_L, k)$ correspond to higher-order mechanisms~; 
in particular, the contribution $(k_S = 1, k_L = 1, k = 0)$ may correspond to
third-order mechanisms taking into account the spin-orbit interaction within
the $3d^7n'\ell'$ configurations. (Contributions of the type 
$(k_S \ne 0, k_L, k)$ were originally introduced in 
Refs.~3 and 4 in the case of lanthanide ions.) 
Alternatively, the contributions $(k_S \ne 0, k_L, k)$ can be 
considered as (minimal) phenomenological extensions of the 
contribution $(k_S = 0, k_L = 2, k = 2)$.$^{8}$ 

Let us now describe the state vectors to be used for calculating the matrix
elements of $H_{eff}$. The initial state vectors can be developed as 
$$
| i \Gamma \gamma ) \; \equiv \;
  | 3d^8 i \Gamma \gamma) \; = \;
  \sum_{SLJ} \;
  | 3d^8 SLJ \Gamma \gamma) \; c(SLJ \Gamma ; i)
\eqno (4)
$$
in terms of symmetry adapted state vectors
$$
| 3d^8 SLJ \Gamma \gamma) \; = \;
  \sum_{M = -J}^J \; | 3d^8 SLJM) \; (JM | J \Gamma \gamma)
\ , \eqno (5)
$$
where the reduction coefficients $(JM | J \Gamma \gamma)$ 
allow to pass from the chain $SO(3) \supset SO(2)$ ($\{ JM             \}$ 
scheme) to 
                   the chain $SO(3) \supset O    $ ($\{ J\Gamma \gamma \}$ 
scheme). The only good quantum numbers in Eq.~(4) 
are $\Gamma$ and $\gamma$, where $\Gamma \equiv \Gamma (O)$ stands
for an irreducible representation class (IRC) of the octahedral group $O$ and 
$\gamma$ is a multiplicity label 
for distinguishing the various partner wave-functions 
associated to the same $\Gamma$. (The label $\gamma$ is really 
necessary only when the dimension of $\Gamma$ is greater
than~1.) Similarly, we take the final state vectors in the form
$$
| f \Gamma ' \gamma ' ) \; \equiv \;
  | 3d^8 f \Gamma ' \gamma') \; = \;
  \sum_{S'L'J'} \; 
  | 3d^8 S'L'J' \Gamma ' \gamma ') \; c(S'L'J'\Gamma ' ; f)
\ . \eqno (6)
$$
The $c$ parameters in (4) and (6) may be obtained by diagonalizing, within the
configuration $3d^8$, some Hamiltonian describing the Coulomb, spin-orbit, and
crystal-field interactions. When dealing with 
Eq.~(9) below, it is important to note  
that the parameters $c(SLJ\Gamma ; i)$ and 
$c(S'L'J'\Gamma' ; f)$ can be chosen in such a way that they do not depend 
on $\gamma$ (for $\Gamma$ fixed) and $\gamma '$ (for $\Gamma '$ fixed), 
respectively.

The matrix element $M_{i(\Gamma \gamma) \rightarrow f(\Gamma ' \gamma ')}$ of
the operator $H_{eff}$ between the state vectors (4) and (6) 
may be calculated by using the Wigner-Racah algebra for
the chain of groups $SO(3) \supset O$. Such a 
calculation has been done in Ref.~8 for a general configuration $n\ell^N$ 
in an arbitrary symmetry $G$. As a result, by taking $G \equiv O$ and $n\ell^N
\equiv 3d^8$, we obtain 
$$
\eqalign{
M_{i(\Gamma \gamma) \rightarrow f(\Gamma ' \gamma ')} \; = \;
& \sum_{S'L'J'} \; \sum_{SLJ} \;
  c(S'L'J' \Gamma ' ; f)^* \; c(S L J  \Gamma   ; i)
\cr
& \sum_{k_Sk_Lk} (-)^{k_S+k_L-k} \;
  C[(k_Sk_L)k] \;
  (3d^8SLJ \Vert W^{(k_Sk_L)k} \Vert 3d^8S'L'J')^*
\cr
& \sum_{\Gamma '' \gamma ''} \;
  f
  \pmatrix{
  J & J' & k\cr
  \Gamma \gamma & \Gamma ' \gamma ' & \Gamma '' \gamma ''\cr
  }^* \;
  \left\{
  {\cal {E}} {\cal {E}}\right \}^{(k)}_{\Gamma '' \gamma ''} \ ,
} \eqno (7)
$$
where
$$
\eqalign{
  f
  \pmatrix{
  J & J' & k\cr
  \Gamma \gamma & \Gamma ' \gamma ' & \Gamma '' \gamma ''\cr
  } \; = \;
  \sum_{MqM'} \; & (-)^{J-M} 
  \pmatrix{
  J & k & J'\cr
  -M & q & M'\cr
  }
\cr
  &(J M | J  \Gamma    \gamma   )^* \; 
   (k q | k  \Gamma '' \gamma '')   \; 
   (J'M'| J' \Gamma '  \gamma ' )   \cr
  }
\eqno (8)
$$
is a coupling coefficient adapted to the chain $SO(3) \supset O$. 

The next step is to calculate the intensity strength
$$
S_{\Gamma \rightarrow \Gamma '} \; = \;
  \sum_{\gamma \gamma '} \;
  \left | M_{i(\Gamma \gamma) \rightarrow f(\Gamma ' \gamma ')} \right |^2
\ . \eqno (9)
$$
The sum on $\gamma$ and $\gamma '$ in (9) can be handled by 
employing~: (i) the
factorization property$^{14}$ for the $f$ coefficients and (ii) the so-called
orthonormality-completeness property$^{15}$ 
for the Clebsch-Gordan coefficients of the group $O$. 
This leads to the final result
$$
  S_{\Gamma \rightarrow \Gamma '} \; = \; 
  \sum_{k=0,2} \; \sum_{\ell = 0,2} \; \sum_{\Gamma ''} \;
  \chi [k \Gamma '' ; \Gamma \Gamma ' ] \;
  \chi [\ell \Gamma '' ; \Gamma \Gamma ' ]^* \;
  \sum_{\gamma ''} \;
  \{ {\cal {E}} {\cal {E}} \}^{(k   )}_{\Gamma '' \gamma ''} \;   \left ( 
  \{ {\cal {E}} {\cal {E}} \}^{(\ell)}_{\Gamma '' \gamma ''} \right )^*
\ , \eqno (10)
$$
which is a particular case of the general result reported in Ref.~11 for a 
configuration $n\ell^N$ in symmetry $G$. The $\chi$ coefficients in (10)
are given by 
$$
\eqalign{
  \chi [K\Gamma '' ; \Gamma \Gamma '] 
\; = \; &[\Gamma '']^{-1/2} \; [\Gamma ]^{1/2} \; 
  \sum_{S'L'J'} \; \sum_{SLJ} \;
  [J]^{-1/2} \cr
 &c(S'L'J'\Gamma ' ; f)^* \;
  c(S L J \Gamma   ; i)   \;
  \sum_{k_Sk_L} \;
  C[(k_Sk_L)K] \cr
 &(-)^{k_S+k_L-K} \;
  (3d^8SLJ \Vert W^{(k_Sk_L)K} \Vert 3d^8S'L'J')^* \;
  (J' \Gamma ' + K \Gamma '' | J \Gamma )\cr
  }
\eqno (11)
$$
with $K = k, \ell$. In Eq.~(11), the coefficient $( \ + \ | \ )$ 
stands for an isoscalar factor$^{14,16}$ 
for the chain $SO(3) \supset O$. Equations (10) and (11) take into
account the fact that the octahedral group $O$ is a multiplicity-free group (so
that no internal multiplicity label $\beta$ is necessary in (10) and (11)).

The polarization dependence in Eq.~(10) is decribed by the symmetry adapted
factors $\{ {\cal {E}} {\cal {E}} \}^{(K)}_{\Gamma '' \gamma ''}$
(with $K = k,\ell )$ defined by 
$$
\{ {\cal {E}} {\cal {E}} \}^{(K)}_{\Gamma ''\gamma ''} \; = \; 
\sum_{Q = -K}^K \; 
\{ {\cal {E}} {\cal {E}} \}^{(K)}_Q \; (KQ | K\Gamma '' \gamma '')
\eqno (12)
$$
in terms of the coupled spherical components
$$
\{ {\cal {E}} {\cal {E}} \}^{(K)}_Q \; = \;
  (-)^{K-Q} \; [K]^{1/2} \; \sum_{x \, y} \; 
  \pmatrix{
  1 & K & 1\cr
  x & -Q & y\cr
  } \; ({\cal {E}})_x \; ({\cal {E}})_y
\ , \eqno (13)
$$
where the spherical components $({\cal {E}})_q$ (with $q = -1, 0, +1)$ of the
polarization vector ${\cal {E}} \equiv {\cal {E}}^{(1)}$ are given by 
$$
({\cal {E}})_0       \; = \; \cos \Phi \ , \qquad \quad \qquad
({\cal {E}})_{\pm 1} \; = \;
  \mp {{1}\over {\sqrt 2}} \; \sin \Phi \; e^{\pm i \theta}
\eqno (14)
$$
for linear polarization and by
$$
\big (
  ({\cal {E}})_{-1}, ({\cal {E}})_0, ({\cal {E}})_{+1} 
\big ) \; = \; (0,0,-1) \quad \hbox {or} \quad (-1,0,0)
\eqno (15)
$$
for circular polarization. We shall continue to develop the formalism for a
general polarization although the experimental results in Ref.~13 are
concerned only with a linear polarization for which the polar angles 
$(\Phi, \theta)$ are $\Phi = \pi/2$ and 
$\theta = 0$ or $\pi/4$~; this will enable us to make some 
theoretical predictions in Sec.~V.

The number of independent $\chi$ parameters in (10) is controlled by the two
following rules. First, we have a group-theoretical rule indicating that the
sum over $\Gamma ''$ in Eq.~(10) is limited by
$$
  \Gamma '' \subset (k)                       \ , \qquad \quad \qquad
  \Gamma '' \subset (\ell)                    \ , \qquad \quad \qquad
  \Gamma '' \subset \Gamma '^* \otimes \Gamma \ ,
\eqno (16)
$$
where $(k)$ and $(\ell )$ refer to IRC's of the group $SO(3)$ and 
$\Gamma '^* \otimes \Gamma$ is the Kronecker product of 
$\Gamma '^* \equiv \Gamma '$ and $\Gamma$. Therefore, the possible 
$\Gamma ''$ in (10) are determined once the range of values for $k$ and $\ell$
as well as the symmetries $\Gamma$ and $\Gamma '$ of the initial and final
states are known. Second, the range of values for $k$ and $\ell$ is partly
fixed by the following model-dependent rule. For identical photons, we have~:
(i) either $k, \ell = 2$ for second-order mechanism 
(corresponding to $k_S = 0$)
or (ii) $k,\ell = 0$ and 2 for second- plus third-order mechanisms
(corresponding to $k_S = 0$ plus $k_S \not = 0$). The two preceeding rules, used
in conjunction with a model for determining the initial and final state
vectors, allow us to restrict the sums on $k$, $\ell$, and $\Gamma ''$ in the basic
intensity formula (1), as we shall see in Secs.~III and IV.
\aa

\centerline {\bf III. APPLICATION}

We now apply the formalism described in Sec.~II to the case where 
$i = {^3A_2(T_2)}$ for the initial state and $f = 
{^3T_2(A_2,E,T_1,T_2})$, ${^1E(E)}$, ${^3T_1(A_1,E,T_1,T_2)}$, 
${^1T_2(T_2)}$, and ${^1A_1(A_1)}$ 
for the final states. We thus have $\Gamma = T_2$
and $\Gamma ' = A_1, A_2, E, T_1, T_2$. 
Since $(0) = A_1$ and $(2) = E \oplus T_2$
in terms of IRC's of $O$, Eq.~(10) can be simplified to give
$$
  S_{\Gamma \rightarrow \Gamma '} \; = \;
 \left | \chi [0A_1 ; \Gamma \Gamma '] \right |^2 \;
 \left | \{ {\cal {E}} {\cal {E}} \}^{(0)}_{A_1} \right |^2 \; + \;
  \sum_{\Gamma ''} \; 
  \left | \chi [2 \Gamma '' ; \Gamma \Gamma ' ] \right |^2 \;
  \sum_{\gamma ''} \;
  \left | \{ {\cal {E}} {\cal {E}} \}^{(2)}_
  {\Gamma '' \gamma ''} \right |^2 \ , 
\eqno (17)
$$
where the sum over $\Gamma ''$ is limited to those IRC's $E$ and $T_2$ occuring
in $\Gamma ' \otimes \Gamma$.

At this stage, the indices of type $\gamma$ (like 
$\gamma''$ in Eq.~(17)) can be taken in the form $\gamma \equiv \Gamma
(D_4)\Gamma(D_2)$, where $\Gamma(D_4)$ and $\Gamma(D_2)$ denote IRC's of the
subgroups $D_4$ and $D_2$ of $O$, respectively. Then, the 
polarization dependence in Eq.~(17) is easily calculated using the
chain $SO (3) \supset O \supset D_4 \supset D_2$ with the 
reduction coefficients $(JM | J \Gamma \gamma)$ defined via$^{15}$ 
$$
\eqalign{
  | 0A_1A_1A  ) \; &= \; | 00) \ , 
\cr
  | 2EA_1A    ) \; &= \; | 20) \ , 
\cr
  | 2EB_1A    ) \; &= \; {{1} \over {\sqrt 2}} \; [ | 22) \; + \; | 2-2)] 
\ , 
\cr
  | 2T_2B_2B_1) \; &= \; {{1} \over {\sqrt 2}} \; [ | 22) \; - \; | 2-2)] 
\ , 
\cr
  | 2T_2EB_2  ) \; &= \; {{i} \over {\sqrt 2}} \; [ | 21) \; - \; | 2-1)] 
\ , 
\cr
  | 2T_2EB_3  ) \; &= \; {{-1}\over {\sqrt 2}} \; [ | 21) \; + \; | 2-1)]
}
\eqno (18)
$$
in terms of symmetry adapted state vectors of type 
$| J \, \Gamma(O) \, \Gamma(D_4) \, \Gamma(D_2))$.

From Eqs.~(12), (13), and (18), we get 
$$\eqalign{
  \{ {\cal {E}} {\cal {E}} \}^{(0)}_{A_1A_1A} \; 
 &= \; {{-1}\over {\sqrt 3}} \qquad \hbox {or} \qquad 0 \ , \cr
  \{ {\cal {E}} {\cal {E}} \}^{(2)}_{EA_1A} \; 
 &= \; {{1}\over {\sqrt 6}}\;
  (3\cos ^2 \Phi - 1) \qquad \hbox {or} \qquad 0 \ , \cr
  \{ {\cal {E}} {\cal {E}} \}^{(2)}_{EB_1A} \; 
 &= \; {{1}\over {\sqrt 2}}\;
  \sin ^2 \Phi \; \cos 2 \theta \qquad \hbox {or} \qquad 
{{1}\over {\sqrt 2}} \ , \cr
  \{ {\cal {E}} {\cal {E}} \}^{(2)}_{T_2B_2B_1} \; 
 &= \; {{i}\over {\sqrt 2}}\;
  \sin^2 \Phi \; \sin 2 \theta 
  \qquad \hbox {or} \qquad \pm {{1}\over {\sqrt 2}} \ , \cr
  \{ {\cal {E}} {\cal {E}} \}^{(2)}_{T_2EB_2} \; 
 &= \; {{-i}\over {\sqrt 2}}\;
  \sin 2 \Phi \; \cos \theta \qquad \hbox {or} \qquad 0 \ , \cr
  \{ {\cal {E}} {\cal {E}} \}^{(2)}_{T_2EB_3} \; 
 &= \; {{i}\over {\sqrt 2}}\;
  \sin 2 \Phi \; \sin \theta \qquad \hbox {or} \qquad 0
  }
  \eqno (19)
$$
for linear or circular polarization, respectively. Thus, 
the possible polarization factors in Eq.~(17) are
$$
\eqalign{
a \; &= \; 
\left | \{ {\cal {E}} {\cal {E}} \}^{(0)}_{A_1} \right |^2 
\; = \; {1 \over 3} 
\qquad {\hbox {or}} \qquad 0 \ , \cr
b \; &= \; \sum_{\gamma''} \;
\left | \{ {\cal {E}} {\cal {E}} \}^{(2)}_{E \gamma ''} \right |^2
\; = \; {1 \over 6} \; 
[(3 \cos^2 \Phi - 1)^2 + 3 \sin^4 \Phi \; \cos^2 2 \theta]
\qquad {\hbox {or}} \qquad {1 \over 2} \ , \cr 
c \; &= \; \sum_{\gamma''} \;
\left | \{ {\cal {E}} {\cal {E}} \}^{(2)}_{T_2 \gamma ''} \right |^2
\; = \; {1 \over 2} \; 
(\sin^4 \Phi \; \sin^2 2 \theta + \sin^2 2 \Phi)
\qquad {\hbox {or}} \qquad {1 \over 2} \ .
} \eqno (20)
$$
Therefore, Eq.~(17) can be expressed as
$$
  S_{T_2 \rightarrow \Gamma '} \; = \;
a \; \left | \chi [0A_1 ; T_2 \Gamma '] \right |^2 \; + \; 
b \; \left | \chi [2E   ; T_2 \Gamma '] \right |^2 \; + \;
c \; \left | \chi [2T_2 ; T_2 \Gamma '] \right |^2 
\ , \eqno (21)
$$
where $\Gamma' = A_1, A_2, E, T_1, T_2$. It should be 
emphasized that the $\chi$ parameters in Eq.~(21) depend not 
only on the symmetry (i.e., $\Gamma'$) of the involved final state but 
also on the corresponding electronic state vectors (cf., Eq.~(6)). 

The experimental situation described in Ref.~13 corresponds to the wave
number ${\vec k}$ along one cube axis so that $\Phi = \pi/2$. Therefore, we
have
$$
\eqalign{
b \; &= \; {1 \over 2} \; 
( \cos^2 2 \theta \; + \; {1 \over 3} ) \quad {\hbox {or}} \quad {1 \over 2}
\ , \cr 
c \; &= \; {1 \over 2} \; 
\sin^2 2 \theta                         \quad {\hbox {or}} \quad {1 \over 2}
}\eqno (22)
$$
and the $\theta$-dependent intensities $S_{T_2 \to \Gamma'}$ 
with $\Gamma' = A_2, T_2, T_1, E, A_1$ are 
$$
\eqalign{
  S_{T_2 \rightarrow A_2} 
\; &= \;  0 \qquad {\hbox {or}} \qquad 0 \ , 
\cr
  S_{T_2 \rightarrow T_2} \; &= \; {1 \over 3} \; r^2 \; + \; {1 \over 
6} \; s^2 \; (1 \; + \; 3 \cos^2 2 \theta) \; + \; {1 \over 2} 
\; t^2 \; \sin^2 2 \theta \qquad {\hbox {or}}
                          \qquad {1 \over 2} \; (s^2 \; + \; t^2) 
\ , \cr
  S_{T_2 \rightarrow T_1} \; &= \; {1 \over 6} \; u^2 \; 
(1 \; + \; 3 \cos^2 2 \theta) \; + \; {1 \over 2} 
\; v^2 \; \sin^2 2 \theta \qquad {\hbox {or}} 
                          \qquad {1 \over 2} \; (u^2 \; + \; v^2) 
\ , \cr
   S_{T_2 \rightarrow E} \; &= \; {1 \over 2} 
\; w^2 \; \sin^2 2 \theta \qquad {\hbox {or}} 
                          \qquad {1 \over 2} \; w^2 \ , 
\cr
  S_{T_2 \rightarrow A_1} 
 \; &= \; {1 \over 2} \; x^2 \; \sin^2 2 \theta 
\qquad {\hbox {or}} \qquad {1 \over 2} \; x^2
}
\eqno (23)
$$
for linear or circular polarization, respectively. In Eq.~(23), 
we have introduced the non-negative parameters 
$$
\eqalign{
  r^2 &= |\chi [0A_1 ; T_2T_2]|^2 \ , \cr
  s^2 &= |\chi [2E ;T_2T_2]|^2    \ , \cr
  t^2 &= |\chi [2T_2 ; T_2T_2]|^2 \ , \cr
  u^2 &= |\chi [2E ; T_2T_1]|^2   \ , \cr
  v^2 &= |\chi [2T_2 ; T_2T_1]|^2 \ , \cr 
  w^2 &= |\chi [2T_2 ; T_2E]|^2   \ , \cr
  x^2 &= |\chi [2T_2 ; T_2A_1]|^2 \ .
}
\eqno (24)
$$
It should be observed that $r= 0$ if we restrict ourselves to 
second-order mechanisms.

Equation (23) gives the detailed polarization dependence of the 
intensity for the various two-photon transitions. For linear 
polarization, Eq.~(23) can be particularized to the special 
case $\theta =0$ and $\theta= \pi/4$ corresponding to the 
experimental results reported in Ref.~13 
(light polarized along the (100) and (110) axes, 
respectively). Then, the ratios 
considered in Ref.~13 for the 
transitions ${^3A_2(T_2)} \to {^3T_2(E,T_1})$ assume the form
$$\eqalign{
  R_1 &=
  {{S_{T_2 \rightarrow T_1} (\theta = 45^\circ)}\over
   {S_{T_2 \rightarrow T_1} (\theta = 0^\circ)}} =
  {1\over 4} \; \left ( 1 + 3 \, {{v^2}\over {u^2}}\right ) \ , \cr
\cr
  R_2 &=
  {{S_{T_2 \rightarrow T_1} (\theta = 45^\circ)}\over
   {S_{T_2 \rightarrow E} (\theta = 45^\circ)}} =
  {1\over 3} \; {{u^2 + 3 \, v^2} \over {w^2}}
}\eqno (25)
$$
and will serve for testing our theory. 
We shall also consider 
ratios, similar to $R_1$, defined for any final state of symmetry
$\Gamma' = A_1, A_2, E, T_1, T_2$ by
$$
R =
  {{S_{T_2 \rightarrow \Gamma'} (\theta = 45^\circ)}\over
   {S_{T_2 \rightarrow \Gamma'} (\theta = 0^\circ)}}
\ , \eqno (26)
$$
where, according to Eqs.~(21) and (22), we have
$$
\eqalign{
S_{T_2 \rightarrow \Gamma '} (\theta = 45^\circ) \; &= \;
{1 \over 3} \; \left | \chi [0A_1 ; T_2 \Gamma '] \right |^2 \; + \; 
{1 \over 6} \; \left | \chi [2E   ; T_2 \Gamma '] \right |^2 \; + \;
{1 \over 2} \; \left | \chi [2T_2 ; T_2 \Gamma '] \right |^2 \ , \cr 
\cr 
S_{T_2 \rightarrow \Gamma '} (\theta = 0^\circ ) \; &= \;
{1 \over 3} \; \left | \chi [0A_1 ; T_2 \Gamma '] \right |^2 \; + \; 
{2 \over 3} \; \left | \chi [2E   ; T_2 \Gamma '] \right |^2 \ . 
} \eqno (27)
$$
Note that in the case where $f = {^3T_2}(\Gamma' = T_1)$, 
$R$ is nothing but $R_1$. 
\aa

\centerline {\bf IV. RESULT}

The intensity parameters $\chi[k \Gamma'' ; T_2 \Gamma']$ in 
Eq.~(24) can be obtained with the help of the definition (11). 
The expansion coefficients $c(S L J \Gamma ; i)$ and 
                           $c(S'L'J'\Gamma'; f)$ in Eq.~(11) 
can be derived by diagonalizing the appropriate 
Hamiltonian within the configuration 
$3d^8$. Such a diagonalization has been done by 
Campochiaro {\it et al.}$^{13}$ who used the simple model 
developed by Liehr and Ballhausen.$^{17}$ However, the 
wave-functions in Ref.~13 are given in the 
strong-field basis and, in order to get the $c$ coefficients, 
they have to be transformed to the weak-field basis. This can be 
achieved by means of the transformation matrices set up in 
Ref.~17. As a result, the wave-functions corresponding to 
the sixteen strong-field states considered in Ref.~13 are 
described in Table I in a weak-field basis. 

Looking at Table I, we note that in most cases the dominating 
weak-field components of the initial state $^3A_2(T_2)$ can be 
connected to the dominating weak-field components of the final 
state via the tensor operator ${\bf W}^{(02)2}$ arising in the 
standard second-order model of intra-configurational two-photon 
absorption. This is an indication that the 
second-order mechanism depicted by $C[(02)2]$ in Eq.~(11) 
should be sufficient to interpret the two-photon transitions in 
the case of Ni${^{2+}}$ in MgO. We may then limit our analysis to 
the contribution ($k_S = 0, k_L = 2, k = 2$). The only free 
parameter is then $C[(02)2]$. It can be calculated from 
$$
C[(02)2] \; = \; 
- {1 \over 5} \; {\sqrt {14 \over 3}} \; e^2 \; 
{ {\big( 3 d | r | 4 p \big) ^2} \over {\left [ \hbar \omega - E(4p) \right ]} }
\eqno (28)
$$
which follows from Eq.~(3) provided we restrict the sum 
over $n' \ell'$ to $n' \ell' \equiv 4p$. The magnitude of the parameter 
$C[(02)2]$ can be estimated by taking the reasonable value 
$E(4p) \approx (3/4) R_{\infty}$ 
($R_{\infty} = {\hbox {Rydberg constant}}$) 
and by using the radial 
integrals tabulated in Ref.~18. However, this 
magnitude is the same for all the transitions to be considered, 
so we can normalize it to any convenient value. For our 
calculations we take $C[(02)2] = 4 \; {\sqrt {35 \over 3}}$. 

The reduced matrix elements in Eq.~(11) follow from the tables 
in Ref.~19 owing to 
${\bf W}^{(02)2} = {\sqrt {5 \over 2}} \; {\bf U}^{2}$. The 
isoscalar factors $(J' \Gamma' + K \Gamma'' \vert J \Gamma)$ in 
(11) can be calculated by applying the method developed in 
Ref.~14 to the data of Ref.~15~; 
alternatively, they can be deduced from the $SO(3) \supset O$ 
factors of Ref.~20 by means of the connecting formula
$$
(J' \Gamma' + K \Gamma'' \vert J \Gamma)
\; = \; 
\sqrt{ {[J] \over [\Gamma]} } \;  
  \pmatrix{
  J      & K         & J'       \cr
  \Gamma & \Gamma '' & \Gamma ' \cr
  }^{SO(3)}_{O} 
\eqno (29)
$$
which arises by expressing, in the notations of Refs.~15 
and 20, the Wigner-Eckart theorem for the 
groups $SO(3)$ and $O$ in an $SO(3) \supset O$ basis. 

By following the scheme just described, we finally obtain the 
values of the $\chi$ parameters, the intensity parameters 
(23), and the intensities (27). These 
intensities together with the ratio $R$ are presented in Table II. 
\aa

\centerline {\bf V. CONCLUSION}
\parskip = 0.35 true cm

We have concentrated in this paper on a model for describing 
two-photon intra-configurational transitions for an ion with 
$d^8$ or $d^2$ configuration in a surrounding of octahedral 
symmetry. The model is based on the consideration of 
second-order mechanisms  
with ionic wave-functions as the initial and 
final state vectors. Furthermore, the 
information on symmetry manifests itself in this model 
through the quantitative 
use of symmetry adaptation techniques for 
the chain of groups $SO(3) \supset O$. The model leads to 
intensity formulas in the spirit of those derived in 
Refs.~8, 9, and 11 for a configuration $n \ell^N$ in an 
arbitrary symmetry. These formulas exhibit the polarization 
dependence for linearly and circularly polarized photons in 
terms of intensity parameters which depend on a single parameter 
(viz., $C[(02)2]$).

In the case of linearly polarized photons, the application of 
the latter formulas to Ni$^{2+}$ in MgO yields theoretical 
intensities in reasonable agreement with the experimental 
values of Ref.~13. There is no experimental result for 
circularly polarized photons, so that our results provide 
predictions in this case.

Our results concern the two-photon transitions from the ground 
state $^3A_2(T_2)$ to the first fifteen excited states of 
MgO:Ni$^{2+}$. The polarization ratios $R_1$ and $R_2$ defined 
by Eq.~(25) for the transitions $^3A_2(T_2)$ $\to$ $^3T_2(E,T_1)$ 
are of special interest for testing purposes because of the 
particularly good resolution of these transitions. The reported 
experimental values of $R_1$ and $R_2$ are 3.0 and 1.1, 
respectively. 
From Table II, we can obtain the theoretical values 
of $R_1$ and $R_2$ : we find $R_1 \equiv R = 0.95$ and we 
calculate, using Eq.~(25), $R_2 = 1.04$. Note that the 
theoretical values obtained in Ref.~13 are 
$R_1 = 220$ and $R_2 = 25$. 
To get the experimental 
values of $R_1$ and $R_2$, the parameters $u^2$, $v^2$, and $w^2$ 
occurring in Eq.~(23) would have to satisfy the relations 
$u^2/w^2 = 0.28$ and $v^2/w^2 = 1$. The introduction of the 
latter relations into Eq.~(23) 
may give predictions on the polarization dependence of the 
transitions $^3A_2(T_2)$ $\to$ $^3T_2(E,T_1)$ for any value 
of $\theta$ in term of a single parameter (say, $w^2$). 

Is it possible to improve the model~? As can be easily proved, 
the third-order (spin-orbit) correction$^{3,4}$ has in our case 
very limited significance. First, its leading term is 
proportional to the tensor operator ${\bf W}^{(11)0}$ and gives a vanishing 
contribution since this operator cannot link $^3A_2(T_2)$ and 
$^3T_2(E,T_1)$. Second, the reamaining terms (arising from the 
tensors ${\bf W}^{(1k_L)2}$) give, according to 
an aside calculation, corrections of an order of a few 
percents.   

Another possible improvement of the model might come, as already 
mentioned by McClure and co-workers,$^{13}$ from dynamic 
contributions of the ligands. In this respect, however, much 
more extensive and involved studies are needed to reach 
conclusions. 

To close, it is worth noticing that the approach presented in 
Secs.~II and III can be extended to any other transition-metal 
ion or any rare-earth ion in an arbitrary symmetry, possibly 
with two different photons (see Ref.~11 
for further details). We also mention that a similar formalism 
can be developed for inter-configurational two-photon 
transitions.$^{21}$ 
\parskip = 0.4 true cm

\vfill\eject

\vsize = 22 true cm
\hsize = 17 true cm
\baselineskip = 0.72 true cm
\hrule
\bigskip

\centerline {
TABLE I. The wave-functions of the ion Ni$^{2+}$ in MgO in a weak-field basis.
}
\bigskip
\hrule

$$\eqalign{
^3A_2(T_2) = & \ 0.02 \ i \ |^1G_4 ) + 0.02 \ i \ |^1D_2 ) - 0.55 \ |^3F_3 ) 
              +0.72 \ i \ |^3F_4 ) - 0.41 \ i \ |^3F_2 )\cr \cr
^3T_2(E) = & - 0.03 \ |^3P_2 ) - 0.475 \ |^3F_2 ) - 0.87 \ |^3F_4)\cr
^3T_2(T_1) = & - 0.03 \ |^3P_1 ) - 0.72 \ |^3F_3 ) + 0.69 \ i \ |^3F_4)\cr
^3T_2(T_2) = & \ 0.03 \ i \ |^3P_2 ) + 0.31 \ |^3F_3 ) + 0.64 \ i \ |^3F_4)
               + 0.70 \ i \ | ^3F_2 )\cr
^3T_2(A_2) = & \ - i \ |^3F_3 )\cr \cr
^1E(E) = & - 0.44 \ |^1G_4 ) + 0.73 \ |^1D_2 ) + 0.25 \ |^3P_2 ) 
           - 0.36 \ |^3F_2 ) + 0.27 \ |^3F_4 )\cr \cr 
a \ ^3T_1(A_1) = & \ 0.34 \ i \ |^3P_0 ) - 0.94 \ i \ |^3F_4 )\cr
a \ ^3T_1(T_1) = & - 0.35 \ |^3P_1 ) + 0.65 \ |^3F_3 ) + 0.66 \ i \ |^3F_4 )\cr
a \ ^3T_1(T_2) = & \ - 0.39 \ i \ |^3P_2 ) + 0.73 \ |^3F_3 ) 
                     + 0.28 \ i \ |^3F_4 ) - 0.49 \ i \ |^3F_2 )\cr
a \ ^3T_1(E) = & \ 0.33 \ |^1G_4 ) - 0.38 \ |^1D_2 ) + 0.33 \ |^3P_2 ) 
                - 0.73 \ |^3F_2 ) + 0.32 \ |^3F_4 )\cr \cr
^1T_2(T_2) = & \ 0.38 \ i \ |^1G_4 ) + 0.82 \ i \ |^1D_2 ) 
              + 0.39 \ i \ |^3P_2 ) + 0.11 \ |^3F_3 )\cr 
             &+ 0.04 \ i \ |^3F_4 ) - 0.07 \ i \ |^3F_2 )\cr \cr
^1A_1(A_1) = & \ 0.25 \ i \ |^1S_0 ) + 0.95 \ i \ |^1G_4 ) - 0.05 \ i \ |^3P_0 )
              - 0.10 \ i \ |^3F_4 )\cr \cr
b \ ^3T_1(E) = & - 0.91 \ |^3P_2 ) - 0.33 \ |^3F_2 ) + 0.24 \ |^3F_4 )\cr
b \ ^3T_1(T_2) = & \ 0.28 \ i \ |^1G_4 ) + 0.32 \ i \ |^1D_2 ) 
                  - 0.83 \ i \ |^3P_2 ) - 0.28 \ |^3F_3 )\cr
                 &- 0.11 \ i \ |^3F_4 ) + 0.19 \ i \ |^3F_2 )\cr
b \ ^3T_1(T_1) = & \ 0.94 \ |^3P_1 ) + 0.21 \ |^3F_3 ) + 0.28 \ i \ |^3F_4)\cr
b \ ^3T_1(A_1) = & \ - 0.94 \ i \ |^3P_0 ) - 0.34 \ i \ |^3F_4 )\cr
}$$

\vfill\eject

\baselineskip = 0.65 true cm
\hrule
\bigskip

\noindent
TABLE II. The results of two-photon intensity calculations. The 
intensities $S$ and the ratio $R$ are defined by Eqs.~(26) and (27). 
To get the intensities defined by (9), each $S$ has to be multiplied 
by $(3/35) \; \{ C[(02)2]/4 \}^2$.
\bigskip
\hrule
\bigskip
\bigskip

\def\init{\tabskip 0pt\offinterlineskip}
\def\crr{\cr\noalign{\hrule}}
\def\crb{\cr\noalign{\hrule height1.5pt}}
$$\vbox{\init\halign to 16.5 true cm{\strut#&\vrule width1.5pt#%
\tabskip=0em plus 2 em&
\hfil$#$\hfil&
\vrule#&
\hfil$#$\hfil&
\vrule#&
\hfil$#$\hfil&
\vrule#&
\hfil$#$\hfil&
\vrule#&
\hfil$#$\hfil&
\vrule width1.5pt#\tabskip 0pt\crb
&& && && && && &\cr
&&\hbox {Strong-field} &&\hbox {Substate} &&\hbox {Intensity} 
&&\hbox {Intensity} && R &\cr
&&\hbox {term} &&\hbox {symmetry} 
&&S_{T_2 \rightarrow \Gamma '}(0^\circ) 
&&S_{T_2 \rightarrow \Gamma '}(45^\circ) && &\cr
&& &&(\Gamma ') && && && &\cr
&& && && && && &\crr
&& && && && && &\cr
&&^3T_2\hfill &&E\hfill &&0\hfill &&1.02\hfill &&- &\cr
&& &&T_1\hfill &&1.11\hfill &&1.06\hfill &&0.95\hfill &\cr
&& &&T_2\hfill &&1.27\hfill &&0.43\hfill &&0.34\hfill &\cr
&& &&A_2\hfill &&0\hfill &&0\hfill &&- &\cr
&& && && && && &\cr
&&^1E\hfill &&E \hfill &&0\hfill &&0.35\hfill &&- &\cr
&& && && && && &\cr
&&a\ ^3T_1\hfill &&A_1 \hfill &&0\hfill &&0.44\hfill &&- &\cr
&& &&T_1\hfill &&3.51\hfill &&0.90\hfill &&0.26\hfill &\cr
&& &&T_2\hfill &&0.92\hfill &&7.30\hfill &&7.93\hfill &\cr
&& &&E \hfill &&0\hfill &&0.93\hfill &&- &\cr
&& && && && && &\cr
&&^1T_2\hfill &&T_2 \hfill &&1.28\hfill &&1.68\hfill &&1.31\hfill &\cr
&& && && && && &\cr
&&^1A_1\hfill &&A_1 \hfill &&0\hfill &&0.01\hfill &&- &\cr
&& && && && && &\cr
&&b\ ^3T_1\hfill &&E\hfill &&0\hfill &&3.24\hfill &&- &\cr
&& &&T_2 \hfill &&5.88\hfill &&7.18\hfill &&1.22\hfill &\cr
&& &&T_1\hfill &&5.51\hfill &&7.50\hfill &&1.36\hfill &\cr
&& &&A_1\hfill &&0\hfill &&0.06\hfill &&- &\cr
&& && && && && &\crb
}}$$

\vfill\eject

\centerline {\bf REFERENCES}

\vsize = 22 true cm
\hsize = 17 true cm
\baselineskip 0.85 true cm

\itemitem{[*]} On leave of absence from the Institute for Low Temperature 
and Structure Research, Polish Academy of Sciences, 50~-~950 
Wroc\l aw, Poland.

\itemitem{[1]} J.D. Axe, Jr., Phys.~Rev. {\bf 136}, A42 (1964).

\itemitem{[2]} T.R. Bader and A. Gold, Phys.~Rev. {\bf 171}, 997 (1968). 

\itemitem{[3]} B.R. Judd and D.R. Pooler, J. Phys. C : Solid State Phys.
{\bf 15}, 591 (1982).

\itemitem{[4]} M.C. Downer and A. Bivas, Phys.~Rev.~B {\bf 28}, 
3677 (1983)~; M.C. Downer, G.W. Burdick and D.K. Sardar, J. Chem. Phys. 
{\bf 89}, 1787 (1988). 

\itemitem{[5]} M.F. Reid and F.S. Richardson, Phys.~Rev.~B 
{\bf 29}, 2830 (1984). 

\itemitem{[6]} J. Sztucki and W. Str\c ek, Phys.~Rev.~B {\bf 34}, 3120 
(1986)~; Chem.~Phys.~Lett. {\bf 125}, 520 (1986).

\itemitem{[7]} K. Jankowski and L. Smentek-Mielczarek, 
Molec.~Phys. {\bf 60}, 1211 (1987)~; L. Smentek-Mielczarek and 
B.A. Hess, Jr., Phys.~Rev.~B {\bf 36}, 1811 (1987).

\itemitem{[8]} M. Kibler and J.C. G\^acon, Croat. Chem. Acta 
{\bf 62}, 783 (1989). 

\itemitem{[9]} J.C. G\^acon, J.F. Marcerou, M. Bouazaoui, B. Jacquier, 
and M. Kibler, Phys.~Rev.~B {\bf 40}, 2070 (1989)~; 
J.C. G\^acon, M. Bouazaoui, B. Jacquier, M. Kibler, L.A. 
Boatner, and M.M. Abraham, Eur. J. Solid State Inorg. Chem. 
{\bf 28}, 113 (1991).

\itemitem{[10]} J. Sztucki and W. Str\c ek, Chem.~Phys. {\bf 143}, 347 (1990).

\itemitem{[11]} M. Kibler, in~: {\it Symmetry and Structural 
Properties of Condensed Matter}, Eds.  W. Florek, T. Lulek 
and M. Mucha (World, Singapore, 1991).  

\itemitem{[12]} R. Moncorg\'e and T. Benyattou, Phys.~Rev.~B 
{\bf 37}, 9186 (1988). 

\itemitem{[13]} C. Campochiaro, D.S. McClure, P. Rabinowitz 
and S. Dougal, Phys.~Rev.~B {\bf 43}, 14 (1991). 

\itemitem{[14]} M. Kibler, C.R. Acad. Sc. (Paris) B {\bf 268}, 1221 
(1969)~; Int. J. Quantum Chem. {\bf 23}, 115 (1983)~; 
Croat. Chem. Acta {\bf 57}, 1075 (1984).  

\itemitem{[15]} M.R. Kibler, in~: {\it Recent Advances in Group 
Theory and Their Application to Spectroscopy}, Ed. J.C. Donini 
(Plenum Press, New York, 1979). 
 
\itemitem{[16]} G. Racah, Phys.~Rev. {\bf 76}, 1352 (1949). 

\itemitem{[17]} A.D. Liehr and C.J. Ballhausen, Ann.~Phys.~(N.Y.) 
{\bf 5}, 134 (1959). 

\itemitem{[18]} F.M.O. Michel-Calendini and M.R. Kibler, 
Theoret. Chim. Acta (Berl.) {\bf 10}, 367 (1968). 

\itemitem{[19]} C.W. Nielson and G.F. Koster, {\it Spectroscopic 
Coefficients for the $p^n$, $d^n$, and $f^n$ Configurations}
(M.I.T. Press, Cambridge, Mass., 1963).

\itemitem{[20]} P.H. Butler, {\it Point Group Symmetry 
Applications. Methods and Tables} (Plenum Press, New York, 
1981). 

\itemitem{[21]} M. Kibler and M. Daoud, preprint LYCEN~9117 
(IPNL, Lyon, 1991). 

\bye